\documentclass[aps,prb,superscriptaddress,twocolumn,showpacs,floats]{revtex4}
\usepackage{graphicx}

\begin{document}

\title{ Systematic generation of finite-range atomic basis sets for 
        linear-scaling calculations}

\author{ Eduardo Anglada }
\affiliation{ Departamento de F\'{\i}sica de la Materia Condensada,
          Universidad Aut\'onoma de Madrid,
          E-28049 Madrid, Spain }
\affiliation{ Institut de Ci\`encia de Materials de Barcelona, CSIC,
          Campus de la UAB, Bellaterra,
          08193 Barcelona, Spain }

\author{ Jos\'e M. Soler }
\affiliation{ Departamento de F\'{\i}sica de la Materia Condensada,
          Universidad Aut\'onoma de Madrid,
          E-28049 Madrid, Spain }

\author{ Javier Junquera}
\affiliation{ Institut de Physique, B\^atiment B5, Universit\'e de Li\`ege,
              B-4000 Sart-Tilman, Belgium }

\author{ Emilio Artacho }
\affiliation{ Department of Earth Sciences, 
          University of Cambridge,
          Downing St., Cambridge CB2 3EQ, United Kingdom }

\date{today}

\begin{abstract}
  Basis sets of atomic orbitals are very efficient for density functional 
calculations but lack a systematic variational convergence.
  We present a variational method to optimize numerical atomic orbitals 
using a single parameter to control their range.
  The efficiency of the basis generation scheme is tested and compared with
other schemes for multiple $\zeta$ basis sets. 
  The scheme shows to be comparable in quality to other widely used schemes 
albeit offering better performance for linear-scaling computations.
\end{abstract}



\maketitle


   The last few years have seen the development of implementations
of the density functional theory (DFT)
\cite{Kohn-Sham1965} in which the computer time and
memory scale linearly with the number $N$ of atoms in the system
studied \cite{Ordejon1998:revTBON,Goedecker1999:RMP}.
   These so-called order-$N$ (O($N$)) methods have increased
considerably the need of accurate and efficient basis sets of
finite range.
   While high accuracy can be achieved with flexible linear
combinations of atomic orbitals (LCAO), high efficiency requires
the orbitals to be as localized as possible.
   Numerical atomic orbitals (NAO's) are well suited to linear
scaling methods because they are very flexible, can be strictly
localized, and few of them are needed for accurate results.
   Their main drawback is the lack of a systematic procedure to ensure a
rapid variational convergence with respect to the number of basis
orbitals and to the range and shape of each orbital.

   In the context of the ab initio pseudopotential method for solids,
an early proposal were the `fireballs' of Sankey and Niklewski:
solutions of the radial Schr\"odinger equation for an isolated
pseudo-atom confined in a spherical hard potential box
\cite{Sankey-Niklewski1989}.
   Subsequent works proposed different recipes to find
multiple-$\zeta$ and polarization orbitals
\cite{Lippert-Hutter-Parrinello1996,Artacho1999,Soler2002}.
   In a recent work \cite{Junquera2001:bases}, a method was proposed 
to optimize the shape of the orbitals by substituting the hard box by 
a soft confining spherical potential
\cite{Porezag1995,Horsfield1997,Junquera2001:bases}.
   This confining potential, which may be different for each
atomic orbital, depends on a series of parameters which determine
the orbital's shape.
   The parameters are then adjusted to minimize the energy of a prototype
 molecule or solid.
   If the confining potentials diverge at given cutoff radii,
the orbitals become strictly zero beyond those radii.
   However, if the cutoff radii themselves are included as
variational parameters, without constraints to impose a small
range, the resulting orbitals tend to become very extended, 
with long tails that generally have no particular significance 
for the condensed system, but which limit severely their efficiency.
   In the present work, we propose a simple procedure to compress
the orbital radii by introducing a fictitious pressure.
   This allows to balance efficiency versus accuracy in a continuous 
and well controlled way.
   In addition, we evaluate the variational completeness of the
resulting orbital shapes, by adding additional degrees of freedom,
and by exploring alternative generation procedures and comparing
their relative merits.


   Our basis orbitals are products of spherical harmonics
times numerical radial functions centered on atoms.
   The quantum chemistry literature typically distinguishes between
core, valence, polarization, and diffuse basis orbitals.
   In our case, core states are eliminated by the use of 
norm-conserving pseudopotentials \cite{Troullier-Martins1991}.
   The explicit description of semicore electrons as valence is 
performed with the same methods described here, but using a pseudopotential 
for which the semicore electrons occupy the ground state and the valence 
electrons occupy the first excited state (with a radial node).
   In previous works we have designed a specific numerical method
for polarization orbitals \cite{Soler2002}, but here we will use
the same methods for valence and polarization orbitals.
   We will not consider diffuse orbitals in this work.

   When several basis orbitals with the same center and angular 
momentum are used to expand the valence states, we follow the 
standard quantum chemical terminology and call them first-$\zeta$ 
orbital, second-$\zeta$ orbital, etc, even though there are no 
$\zeta$ exponent coefficients in our orbitals.
   We use a different method to generate the first-$\zeta$ orbitals
than that for the subsequent-$\zeta$ orbitals.
   For the first-$\zeta$ orbitals we solve the radial Schr\"odinger
equation for a potential given by the sum of
the full (screened) nonlocal pseudopotential corresponding to 
the angular momentum of the orbital, and a confining potential
of the form 
$V(r) = V_{\rm o} \exp\left[ -(r_c - r_i) / (r - r_i) \right] / (r_c -r) $
which depends on three parameters $r_i$, $V_0$, and the cutoff
radius $r_c$. 
   These parameters are different for each basis orbital and
define its range as well as its shape by allowing a depression of the tail.
   Other confinement schemes have been proposed \cite{Sankey-Niklewski1989,
Porezag1995,Horsfield1997} and are compared with this one in Ref.
[\onlinecite{Junquera2001:bases}].
   To generate the second and subsequent-$\zeta$ orbitals we will
use and compare two possible methods.
   The first one is based on the concept of chemical hardness (CH)
and defines the different-$\zeta$ orbitals as the derivatives of the
ground-state wavefunction of the potential (pseudo plus confining)
with respect to the charge of the atom \cite{Lippert-Hutter-Parrinello1996}.
   In this scheme, there are no independent parameters to fix the 
shape of the higher-than-first-$\zeta$ orbitals.

   The second scheme to generate higher-$\zeta$ orbitals was inspired
by the ``split valence'' (SV) method which is standard in quantum 
chemistry, where orbitals are given by fixed linear combination of gaussians
\cite{Huzinaga1984}.
   The second-$\zeta$ (or triple etc) orbitals are obtained by ``splitting''
the slowest-decaying gaussian(s) to act as independent basis orbital(s). 
   The SV was adapted to numerical atomic orbitals by constructing a 
double-$\zeta$ orbital as one that reproduces the tail of the 
first-$\zeta$ from a matching radius outwards, and runs smoothly inwards 
\cite{Artacho1999,Junquera2001:bases,Soler2002}. 
   Higher-$\zeta$ orbitals are obtained 
repeating the procedure at different radii.

   A variational optimization of a basis set constructed as described 
above can give orbitals with too long cutoff radii $r_c$. 
   In order to reduce
their range in a systematic way we introduce a parameter $P$ with
dimensions of pressure (that we will call ``pressure'' henceforth) and
minimize the ``enthalpy'' $E+PV$, where $E$ is the total energy of some
reference system and $V=(4\pi/3)\sum_{\mu}r^3_{c\mu}$ is the sum
of the volumes of the basis orbitals $\phi_{\mu}$. 
   The convergence of calculated properties with respect to orbital 
range is thus controlled by a single parameter, much in the same way as 
the planewave cutoff controls the convergence of a plane wave basis set.
 
   The reference system for which $E+PV$ is minimized is
a molecule or solid in which the atoms considered have a 
prominent role, and which is small enough to allow many 
selfconsistent calculations with different basis parameters.
   The derivatives of $E$ with respect to those
parameters are generally not available, and we use the 
downhill-simplex method\cite{NumericalRecipes} to minimize it.
   The basis orbitals depend on the described parameters in a 
non-linear way, and several local minima are found in many cases. 
   This is to be expected because different combinations of 
parameters can produce approximately the same optimal shape. 
   Since our parameters have no special physical significance, 
any low local minimum is in principle equally acceptable,  even
though the multiple minima produce a somewhat unpleasant ``noise'' 
in the results reported below.

\begin{figure}[t!]
\includegraphics[clip,bb= 9 9 210 301]{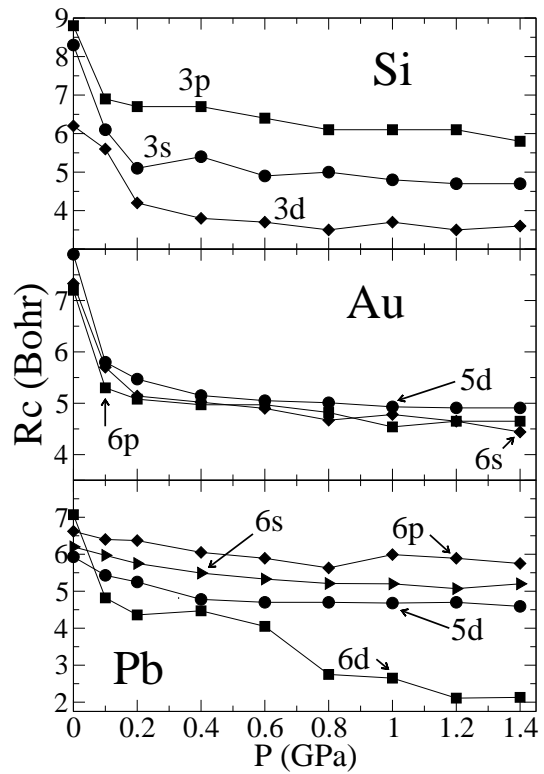}
\caption{
   Cutoff radii of the first-$\zeta$ basis orbitals 
of Si, Au and Pb, as a function of the fictitious pressure parameter $P$
}
\label{Si-Au-Pb}
\end{figure}

   Fig. \ref{Si-Au-Pb},
shows the cutoff radii of the first-$\zeta$ orbitals of
Si, Au, and Pb as a function of the pressure parameter $P$.
   The basis optimizations were performed in their
corresponding bulk solids, with a so-called double-$\zeta$
polarized (DZP) basis set:
   in Si there are double-$\zeta$ $s$ and $p$ shells
and single-$\zeta$ $d$ orbitals;
   in Au there are double-$\zeta$ $s$ and $d$, and single-$\zeta$ $p$; 
   in Pb the $5d$ semicore electrons are included 
in the valence as double-$\zeta$, as are the $s$ and $p$ shells, 
while the $6$d have a single-$\zeta$.
   The second-$\zeta$ orbitals were generated with the SV scheme.
   Polarization orbitals are obtained in the same manner as the other
atomic orbitals.

\begin{figure}[t!]
\includegraphics[clip,bb= 6 0 200 137]{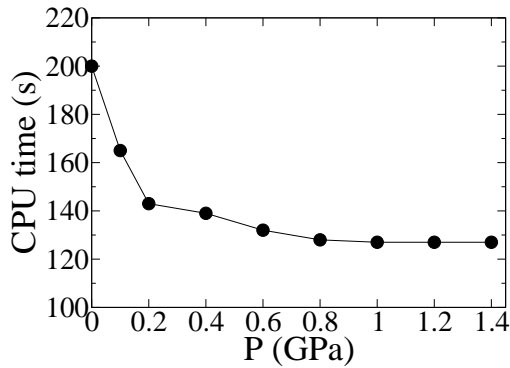}
\caption{
   CPU time for a selfconsistency step of bulk Si (16 atoms per cell) versus
the fictitious pressure $P$ used to compress the cutoff radii
of the DZP basis orbitals.
}
\label{cpu}
\end{figure}

   To give an idea of how the orbital radii affect the basis
efficiency, Fig. \ref{cpu} shows the CPU time required for
the calculation of a selfconsistent step of bulk silicon, as a 
function of the pressure $P$ used to generate the basis.
   The accuracy of the results, as the
orbitals contract, is addressed in Table \ref{aBEc}, 
which shows the variation in lattice parameter, bulk modulus, 
and cohesive energy with $P$.
   The results were obtained using the
{\sc Siesta} method \cite{SanchezPortal1997,Soler2002},
with a well converged real-space integration grid.
   They are compared to experiment and to well-converged plane 
wave calculations, performed with a specific program designed 
to use exactly the same pseudopotential
\cite{Troullier-Martins1991,Kleinman-Bylander1982}, 
exchange correlation functional \cite{Perdew-Zunger1981}, 
and $k$-grid sampling \cite{Monkhorst-Pack1976} 
used in {\sc Siesta}.
   The cohesive energy is calculated 
as the difference between the bulk total energy per atom
(with the chosen basis set) and an atomic calculation in
which the radial Schr\"odinger equation is solved numerically,
without any constraint to the shape or range of the orbitals.
   With this definition the cohesive energy carries the variational
character of the total energy (higher binding energies for better
basis sets).	

\begin{table}
\caption{
   Comparison of structural properties of different systems
as a function of the pressure parameter $P$ (in GPa)
used to generate their basis sets.
   Lattice parameters $a$ in \protect{\AA}, bulk moduli $B$ in GPa and 
cohesive energies $E_c$ in eV.
   The bulk moduli were obtained by fitting the total energy
with a Murnaghan equation of state \cite{Murnaghan1944}.
   A double-$\zeta$ plus polarization basis was used in all cases.
In Pb semicore states where also used.
}
\begin{tabular}{cccccccccc}
\hline \hline
&   &  Exp&  PW &  P=0& 0.2& 0.4&  0.8&  1.2& 1.4 \\
\hline
Si& $a$& 5.43& 5.38& 5.40&  5.38&  5.38&    5.37&   5.36& 5.35 \\
  & $B$&   99&   96&   97&    98&   100&     103&    107& 108 \\
& $E_c$& 4.63& 5.40&  5.36&  5.30&  5.25&  5.12& 4.99&  4.94 \\
\hline
Au& $a$& 4.08& 4.05&  4.06&  4.06&  4.05&    4.02&   4.02&  4.00 \\
  & $B$&  195&  198&   206&   210&   211&    220&     239&  242  \\
& $E_c$&  4.13& 4.36&  4.04&  3.96&  3.95&    3.80&   3.77& 3.66  \\
\hline
Pb& $a$& 4.95& 4.88&  4.90&  4.87&  4.83&   4.79&    4.81&  4.80 \\
  & $B$&   43&   54&    54&    60&    64&     71&      70&  75  \\
& $E_c$& 2.04& 3.77&  3.68&  3.63&  3.48&  3.37&    3.32&  3.29 \\
\hline
MgO& $a$& 4.21& 4.10&  4.11&  4.10&  4.10&    4.11&   4.09&  4.06 \\
   & $B$&  152&  164&  182&   205&   209&      205&   214&   230 \\
 & $E_c$& 10.30& 12.39& 12.18& 12.10& 12.00& 11.86&  11.92& 11.66\\
\hline \hline
\label{aBEc}
\end{tabular}
\end{table}

   It can be seen that a moderate pressure of $\sim 0.2$ GPa
produces a drastic reduction of the orbital radii, with a
correspondingly large reduction of CPU time, without a 
significative change in the results.
   Larger pressures produce additional, though more moderate
gains in basis efficiency, but at the expense of nonneglegible
changes in the results. That small pressure of 0.2 GPa seems to be a
threshold up to which only the very low, not significant, tails are
removed.

\begin{figure}[!t]
\includegraphics[clip,bb=0 5 211 300]{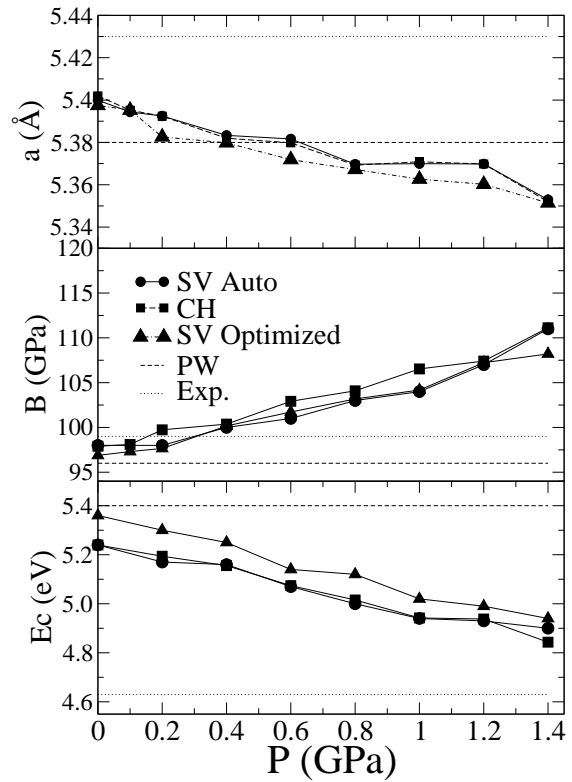}

\caption{
   Equilibrium lattice constant ($a$), bulk modulus ($B$) and cohesive
energy ($Ec$) of bulk Silicon as a 
function of the fictitious pressure parameter $P$. 
   A double-$\zeta$ plus polarization basis was used.
   The second-$\zeta$ orbitals were generated using the
chemical-hardness (CH) and split-valence (SV) schemes.
   For the latter, results are shown for orbitals whose inner
matching radii were generated with a constant split-norm 
parameter of 0.15 or optimized variationally for each value of $P$
(what resulted in the split-norm parameters shown in Fig. 
\ref{splitnorm}).
}
\label{A-B-Ec-Si}
\end{figure}

\begin{figure}[!t]
\includegraphics[clip,bb=5 0 200 140]{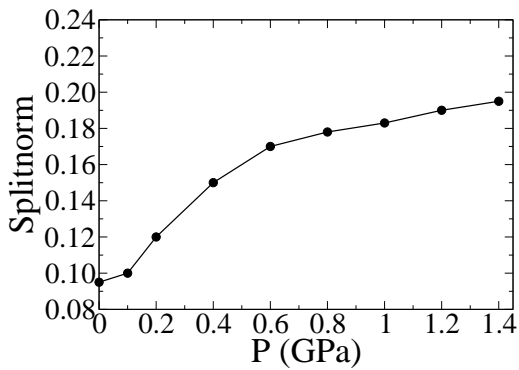}
\caption{
   Optimal value of the split-norm parameter, which determines 
the inner matching radius of the second-$\zeta$ orbitals of silicon
generated with the split-valence scheme.
}
\label{splitnorm}
\end{figure}

\begin{figure}[!t]
\includegraphics[clip,bb= 7 1 211 138]{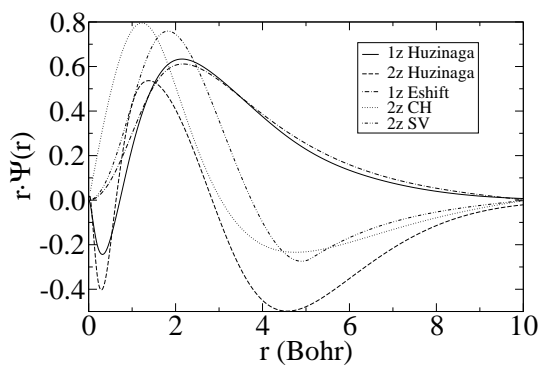}
\caption{
   Radial shape of the first and second $\zeta$ $p$-orbitals of Si. 
   The second-$\zeta$ orbital was generated using the chemical-hardness 
(CH) and split-valence (SV) approaches described in the text. 
   In addition, we show the second-$\zeta$ gaussian orbital
of Huzinaga \cite{Huzinaga1984}.
   The second-$\zeta$ orbitals have been orthogonalized to the 
first-$\zeta$ one to facilitate the comparison.
} 
\label{shapes}
\end{figure}

   The relative merits of the SV and CH methods to generate the
second-$\zeta$ orbitals are considered in Fig. \ref{A-B-Ec-Si}.
   For the SV case, two curves are plotted.
   In one of them, the inner matching radius of the sencond-$\zeta$
orbitals is optimized for every value of $P$.
   In the other one, it is determined by a standard automatic
criterion \cite{Soler2002}, by which the norm of the first-$\zeta$
orbital beyond the matching radious has to be equal to a given
``split-norm'' parameter value of 0.15.
   Fig. \ref{splitnorm} shows the optimized value of this 
parameter, which does not differ much from the standard value.
   As a consequence, it is not surprising that Fig. \ref{A-B-Ec-Si}
 shows a similar quality of the results
using the optimized and standard values.
   The quality is also similar for the CH method, which does
not depend on any variational parameter.
   Again, this is not surprising, in view of the similarity of 
the resulting shapes of the second-$\zeta$ orbitals, which are 
compared in Fig. \ref{shapes} to our SV orbitals and to 
a typical quantum-chemistry gaussian-based polarization orbital
\cite{Huzinaga1984}.
   We may then conclude that the different generating schemes of
second-$\zeta$ orbitals compared here yield basis sets of similar
quality. 	
   Our SV scheme, however, offers higher efficiency for linear-scaling
computations since the range of the higher-$\zeta$ orbitals may be 
restricted to their inner matching radius, without any reduction of the
variational freedom \cite{Soler2002}.

   Finally, we explore to what extent the orbital shapes generated with the
described schemes differ from optimal. 
   To this end, we have added spherical Bessel functions to our generated 
orbitals, not as additional basis functions but to change the shape of the
orbitals in a DZP basis, introducing the coefficients of the linear combination
as the parameters to be optimized.
   Table \ref{bessels} shows the effect in the total energy for bulk silicon.
as subsequent Bessel functions are added to optimize different 
orbitals. The energy reduction is quite moderate,
and considerably smaller than that obtained by introducing additional 
basis orbitals.
   This is true even in the case of the higher-$\zeta$ orbitals,
whose shape depends on just one parameter. 
   It can be thus concluded that the radial shapes of the basis
orbitals are indeed well optimized by the variational freedom
contained in the confining potential, and by the physically motivated
schemes used to generate the higher-$\zeta$ orbitals.

\begin{table}[!t]
\begin{tabular}{lc}
\hline \hline
Basis Size                      & $\Delta E$ (meV) \\
\hline
DZP not optimized               & 230 \\
DZP optimized                   &  40 \\
DZP 4 Bessels in first  $\zeta$ &  33 \\
DZP 4 Bessels in second $\zeta$ &  33 \\
DZP + F                         &  22 \\
DZ2P + F                        &  16 \\
\hline \hline
\end{tabular}
\caption{
   Test of the quality of the DZP optimized basis set of silicon.
   Second-$\zeta$ orbitals were generated with the split-valence method.
   The energies $\Delta E$ are per atom and relative to the 
converged plane wave result. 
   The F stands for the addition of a $f$ angular momentum shell. 
   The 2 in the DZ2P denotes the addition of a second $\zeta$ to the
$d$ polarization orbital.
   The non-optimized basis was obtained with a hard potential 
\cite{Sankey-Niklewski1989} (the radii are as long as in the DZP optimized 
case) and a standard split-norm parameter of 0.15. 
   A zero pressure parameter $P$ was used in all the cases.
}
\label{bessels}
\end{table}

   In conclusion, we have developed a systematic method to construct
accurate and efficient atomic basis orbitals for linear-scaling DFT
calculations.
   The range of the basis sets is controlled by a single parameter, 
that allows to monitor their convergence with range in a simple and systematic way.
   By comparing different generation schemes, and by studying the
effect of additional variational freedom, we have found that our
method produces nearly optimal shapes in multiple-$\zeta$ polarized basis sets.


   We thank Pablo Ordej\'on, Alberto Garc\'{\i}a, Daniel S\'anchez-Portal, 
and Eduardo Hern\'andez for useful discussions. This work was supported by the 
Fundaci\'on Ram\'on Areces and by Spain's MCyT grant BFM2000-1312.

\bibliographystyle{apsrev}

\end{document}